\documentclass[twocolumn]{aastex631}

\usepackage{amssymb}
\usepackage[tbtags]{amsmath}
\usepackage[hang,flushmargin]{footmisc}
\usepackage{multirow}

\newcommand{\gv}[1]{\ensuremath{\mbox{\boldmath$ #1 $}}}

\newcommand{\dsct}[1]{$\delta$~Sct}
\newcommand{\RNum}[1]{\uppercase\expandafter{\romannumeral #1\relax}}
\newcommand{\beq}{\begin{equation}}
\newcommand{\eeq}{\end{equation}}
\newcommand{\bea}{\begin{eqnarray}}
\newcommand{\eea}{\end{eqnarray}}

\defcitealias{Mourabit:23}{MW23}

\shorttitle{Mode coupling in Delta Scutis}
\shortauthors{}

\begin{document}

\title{\bf \large Stability and Dynamics of Three-Mode Coupling in $\delta$ Scuti Stars}

\shortauthors{Mourabit \& Weinberg}
\shorttitle{Three-Mode Coupling in $\delta$ Scuti Stars}

\correspondingauthor{Mohammed Mourabit}
\email{mohammed.mourabit@uta.edu}

\author[0000-0001-5083-7480]{Mohammed Mourabit}
\affiliation{Department of Physics, University of Texas at Arlington, Arlington, TX 76019, USA}

\author[0000-0001-9194-2084]{Nevin N. Weinberg}
\affiliation{Department of Physics, University of Texas at Arlington, Arlington, TX 76019, USA}

\begin{abstract}
Recent observations of $\delta$ Scuti stars find evidence of nonlinear three-mode coupling in their oscillation spectra. 
There are two types of three-mode coupling likely to be important in $\delta$ Scuti stars: (i) direct coupling, in which two linearly unstable  modes (driven by the $\kappa$-mechanism) excite a linearly stable mode, and (ii) parametric coupling, in which one linearly unstable mode excites two linearly stable modes. \citet{Breger:14} find especially strong evidence of direct coupling in the  $\delta$ Scuti star KIC 8054146. However, direct coupling is inherently unstable and cannot be the mechanism by which the  modes saturate and achieve nonlinear equilibrium. By integrating the amplitude equations of small mode networks, we show that the modes can achieve equilibrium if  parametric coupling operates in tandem with direct coupling.  Using mode parameters calculated from a $\delta$ Scuti model, we also find that parametric and direct coupling are likely to be simultaneously active. Importantly, parametric coupling does not necessarily disrupt the  correlations found in KIC 8054146 between the amplitudes and phases of the directly coupled modes. We conclude that $\delta$ Scuti stars are likely impacted by both parametric and direct coupling  and that accounting for both in future large mode network calculations may help explain the complicated mode dynamics observed in many $\delta$ Scuti stars.
\end{abstract}

\section{\bf I\lowercase{ntroduction}}

The primary source of mode driving in $\delta$ Scuti (hereafter \dsct{}) stars is the $\kappa$-mechanism (\citealt{Cox:1963, Chevalier:1971, Pamyatnykh:1999}; for reviews of \dsct{} stars see, e.g., \citealt{Breger:1979, Breger:00, Handler:2009, Guzik:21}).  The modes driven by the $\kappa$-mechanism are linearly unstable and can grow to such large  amplitudes that they excite secondary modes through weakly nonlinear three-mode interactions \citep{Dziembowski:85, Dziembowski:88, Buchler:97, Mourabit:23}. These secondary modes can themselves grow to such large amplitudes that they excite yet more modes, and so on.  If the rate of energy transfer from the primary modes to the secondary modes is sufficient, the growth can saturate and the system will settle into a nonlinear equilibrium. It is unclear, however, whether the rate of energy transfer via weakly nonlinear interactions is sufficient within \dsct{} stars; for example, higher-order nonlinear effects (beyond three-mode interactions) might be needed to saturate the growth, as is  the case in large amplitude pulsators such as Cepheids and RR Lyrae stars (see, e.g., \citealt{Gautschy:1995, Smolec:2008, Buchler:2009, DeSomma:2020}).

In addition to these theoretical considerations,  multiple lines of observational evidence suggest that nonlinear mode interactions shape and modify the oscillation spectra of \dsct{} stars.  For a more detailed discussion of these, we refer the reader to the introduction of \citeauthor{Mourabit:23} (2023; hereafter MW23). Briefly, the evidence includes the observation of mode periods that change much faster than evolutionary models predict \citep{Rodriguez:95, Breger:98, Rodriguez:01, Bowman:21} and of mode amplitudes that vary significantly on time scales of years in many \emph{Kepler} \dsct{} stars \citep{Bowman:16}. Rapid variations are indicative of nonlinear mode coupling because at large  amplitudes the mode dynamics tend to be characterized by short-period limit cycles of growth and decay. In addition to these time-dependent indicators, \citet{Balona:2024} find that the overall shape and richness of oscillation spectra from \emph{TESS} \dsct{} stars shows little, if any, correlation with the locations of the stars in the Hertzsprung-Russell diagram (see also \citealt{Balona:2015}). This contrasts with linear pulsation models which say that similar stars should have similar oscillation spectra. Lastly, \citet{Breger:14} find detailed evidence of mode coupling in the Kepler \dsct{} star KIC 8054146. From the oscillation spectrum, they identify several mode triplets whose frequencies are resonant (the sum of two mode frequencies very nearly equals that of a third mode) and whose mode amplitudes and phases vary with time in a correlated way that is in excellent agreement with the theory of three-mode coupling.

The form of three-mode coupling detected in KIC 8054146 is of a particular type called direct coupling. It involves two linearly unstable modes resonantly exciting a linearly stable mode; we will refer to the former modes as parents and the latter mode as a daughter.   In the context of \dsct{} stars, the parents are driven by the $\kappa$-mechanism while the daughter is not (it is linearly damped rather than driven). Although it might seem that the  daughter could stop the parents' linear growth,  \citet{Dziembowski:82} showed that the direct coupling of three distinct modes is inherently unstable.  A different nonlinear mechanism must therefore be responsible for stopping the parents' growth, and it must do so without disrupting the direct coupling features observed in KIC 8054146.  We will show that another type of three-mode coupling called parametric coupling can satisfy both requirements. In parametric coupling, one parent resonantly excites two daughters.  In \dsct{} stars the two parents that directly couple to a daughter are likely to also each parametrically couple to additional daughters, and thus both types of three-mode coupling are likely to be dynamically active.
     
The paper is organized as follows.  In Section~\ref{sec:NLMC}, we present the formalism we use to study nonlinear mode coupling and in Section~\ref{sec: Three mode coupling} we describe the different types of three-mode coupling likely to be important in \dsct{} stars. In Section~\ref{sec:stability}, we investigate the stability of direct coupling alone and in
Section~\ref{sec:direct + parametric} we investigate the stability of direct coupling acting in tandem with parametric coupling.  In both sections, we consider small mode networks  with parameter values set by hand in order to explore their influence on a system's stability and dynamics. In Section~\ref{sec:delta scuti}, we  construct small mode networks using realistic parameter values calculated from a model of a \dsct{} star and investigate their dynamics. We summarize and conclude in Section~\ref{sec:conclusions}.

\section{\bf N\lowercase{onlinear} 
A\lowercase{mplitude} E\lowercase{quations}}
\label{sec:NLMC}

We study the weakly nonlinear response of a star to a fluid displacement relative to the spherical background using the methodology described in \citetalias{Mourabit:23} and references therein.  Briefly summarizing the approach, let $\gv{\xi}(\gv{x}, t)$ be the Lagrangian displacement field of a fluid element at position $\gv{x}$ and time $t$.  To lowest nonlinear order, the displacement satisfies the equation of motion 
\begin{equation}
\rho \ddot{\gv{\xi}} = \gv{f}_1 [\gv{\xi}] + \gv{f_2} [\gv{\xi}, \gv{\xi}],
\label{eq:eom}
\end{equation}
where $\rho$ is the background density and $\gv{f}_1$ and $\gv{f}_2$ are the linear order and second order nonlinear forces, respectively.  The displacement and its time derivative can be written in terms of  a phase space expansion \citep{Schenk:02}
\begin{align}
    \begin{bmatrix}
    \gv{\xi}(\gv{x}, t)\\
    \gv{\dot{\xi}}(\gv{x}, t)
    \end{bmatrix}= \sum_a q_a(t)\begin{bmatrix}
    \gv{\xi}_a(\gv{x})\\
    -i\omega_a\gv{\xi}_a(\gv{x})
    \end{bmatrix},
\label{eq:xi_expansion}
\end{align}
where the index $a$ labels a linear eigenmode of the star with eigenfunction $\gv{\xi}_a(\gv{x})$, eigenfrequnecy $\omega_a$, and dimensionless amplitude $q_a(t)$. The sum runs over all mode quantum numbers (i.e., radial order $n_a$, angular degree $l_a$, and azimuthal number $m_a$), and frequency sign ($\pm \omega_a$) to allow both a mode and its complex conjugate. We normalize the eigenmodes as
\begin{align}
    2 \omega_a^2 \int \mathrm{d}^3x \, \rho |\gv{\xi_a}|^2 = E_\star,
\end{align}
where $E_\star = GM^2/R$ is a characteristic energy scale of a star with mass $M$ and radius $R$; with this choice of normalization, a mode with amplitude $q_a$ has energy $E_a=|q_a|^2 E_\star$.

Plugging Equation~(\ref{eq:xi_expansion}) into Equation~(\ref{eq:eom}) and using the orthogonality of eigenmodes gives a set of coupled,  time-dependent, nonlinear amplitude equations, which for each mode is given by
\begin{equation}
\dot{q}_a+\left(i\omega_a+\gamma_a\right)q_a = i\omega_a \sum_b\sum_c \kappa_{abc} q_b^\ast q_c^\ast,
\label{eq:amp eqn}
\end{equation}
where the asterisks denote complex conjugation.  The left hand side of Equation~(\ref{eq:amp eqn}) describes an uncoupled harmonic oscillator while the right hand side describes the nonlinear forcing of mode $a$ due to three-mode coupling.  We added a term $\gamma_a q_a$ on the left hand side to describe the linear damping (if $\gamma_a>0$) or linear driving (if $\gamma_a<0$) of a mode.  In \dsct{} stars, the linear damping has contributions from radiative and turbulent dissipation (see Section 3.2 of \citetalias{Mourabit:23}), while the linear driving is due to the $\kappa$-mechanism. 

The three-mode coupling coefficient is a dimensionless quantity given by 
\begin{align}
\kappa_{abc}= \frac{1}{E_\star} \int \mathrm{d}^3x\,  \gv{\xi}_a \cdot \gv{f}_2\left[\gv{\xi}_b,\gv{\xi}_c\right].
\end{align}
Given a triplet of eigenmodes, $\kappa_{abc}$ can be computed using  expression A55-A62 in \cite{Weinberg:12}. The modes couple ($\kappa_{abc} \neq 0$) only if their angular quantum numbers satisfy the angular selection rules $|l_b - l_c| \le l_a \le l_b + l_c$ with $l_a+l_b+l_c$ even and $m_a+m_b+m_c=0$.  It can be shown that $\kappa_{abc}$ is symmetric in all permutations of the mode indices.  Given our choice of mode normalization, we find that for a triplet of low-order modes (i.e., wavelengths of order $R$) within a \dsct{} star, $\kappa_{abc}$ is at most a number of order unity.

\section{\bf T\lowercase{ypes} \lowercase{of}  T\lowercase{hree}-M\lowercase{ode} C\lowercase{oupling}}
\label{sec: Three mode coupling}

Any triplet of modes that satisfies the angular selection criteria will experience some degree of nonlinear coupling and thereby be subject to nonlinear corrections to their mode dynamics.  In practice, however, the nonlinear terms on the right hand side of Equation~(\ref{eq:amp eqn}) only have a significant effect if there is a driving mechanism that excites the modes to large enough amplitudes. In this study, we consider the case where at least one of the modes of each triplet is linearly driven (via the $\kappa$ mechanism).   

We focus on two types of three-mode coupling between the linearly driven modes (parents) and the stable modes they excite (daughters): (1) direct coupling, in which two  parents excite a daughter, and (2) parametric coupling, in which one parent excites two daughters. In one of the first detailed studies of this problem,  \citet{Dziembowski:82} showed that both types of coupling can be important in stars. Whether they are important depends on the magnitudes of the linear driving and damping rates, the coupling coefficient $\kappa_{abc}$, and the detuning $\Delta = \omega_a+\omega_b+\omega_c$ (the frequencies can be positive or negative in our analysis).  The modes form a resonant triplet if $|\Delta|$ is small compared to the magnitude of the individual mode frequencies.

In \citetalias{Mourabit:23} we focused on direct coupling in \dsct{} stars, motivated by the detection of resonantly interacting mode triplets in the \dsct{} star KIC 8054146 \citep{Breger:14}. Over the course of three years, \citet{Breger:14} found that the interacting modes in KIC 8054146  underwent variations in amplitude consistent with two parents exciting a daughter, i.e., direct coupling. As we will show in Section~\ref{sec:delta scuti}, parametric coupling of one parent to pairs of daughters is also likely to be dynamically important in \dsct{} stars.  

Although we do not consider it here, it is possible to have a triplet in which all three modes are linearly driven.  Of course in that case there must be additional modes that the three modes couple to in order to prevent their runaway growth. We also do not consider the case in which the daughters reach such large amplitudes that they nonlinearly excite  additional modes (i.e., granddaughters).  Understanding the influence of these types of interactions requires a study of how a set of coupled modes, some of which are linearly driven, arrive at a nonlinear equilibrium.  This is a complicated problem that has only been studied in detail for a few types of specific systems: for example, $r$-modes in rotating neutron stars \citep{Brink:2004, Brink:2005}, tidally exited $g$-modes in stars hosting hot Jupiters \citep{Essick:2016, Weinberg:2024}, and mixed modes excited by convection in red giants \citep{Weinberg:21}.  Here, as a first step towards  carrying out such a study for \dsct{} stars, we limit ourselves to the simpler problem of  linearly driven parents that excite daughters through a combination of direct and parametric coupling.

\section{\bf S\lowercase{tability} \lowercase{of} D\lowercase{irect} C\lowercase{oupling} }
\label{sec:stability}

\begin{figure}[t!]
\centering
\includegraphics[width=3.4in]{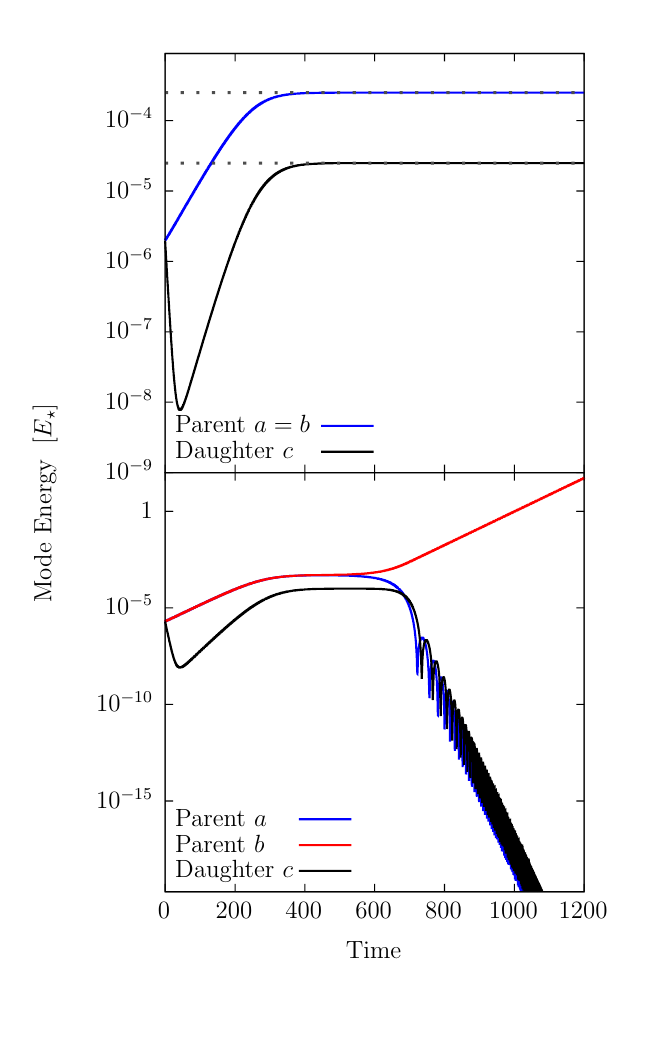}
\caption{Mode energy as a function of time for direct three-mode coupling of a self-coupled parent (mode $a=b$; top panel) or distinct parents ($a\neq b$; bottom panel) to a daughter (mode $c$). In the top panel, the mode frequencies are $\{-0.5, -0.5, 1.0\}$ and the linear driving or damping rates  are $\{-0.001, -0.001, 0.01\}$ while in the bottom they are $\{-0.5, -0.502, 1.0\}$ and $\{-0.0011, -0.001, 0.01\}$, where we list the parents' values before the daughter's. In both panels, $\kappa_{abc}=1.0$. The dotted lines in the top panel show the values of $E_{a,{\rm eq}}$ and $E_{c,{\rm eq}}$ computed using Equation~(\ref{eq:Ea_eq_DC}).
\label{fig:AE_Simple_DC_double}}
\end{figure}

\begin{figure*}[t!]
\centering
\includegraphics[width=7.5in]{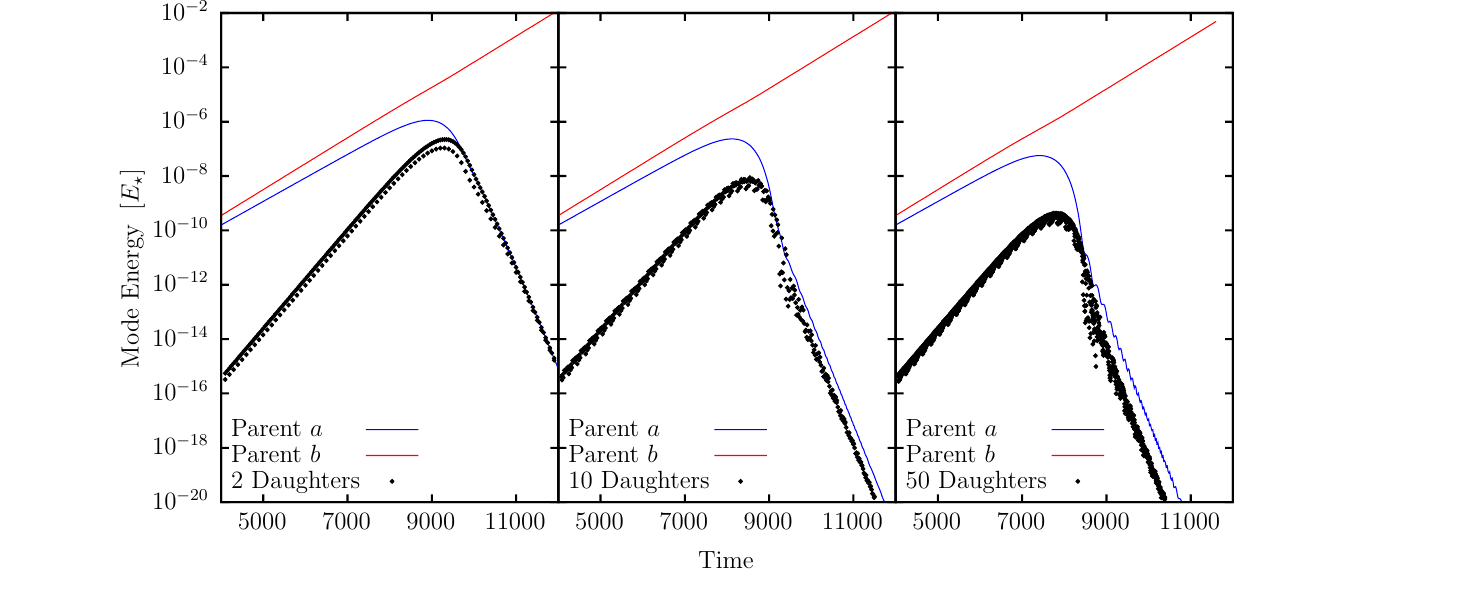}
\caption{Mode energy as a function of time for direct coupling of two distinct parents and $N$ distinct daughters, with $N=2$ (left panel), $N=10$ (middle panel), and $N=50$ (right panel). The parent frequencies and driving rates are $\omega_a=\omega_b=-0.5$ and  $\gamma_a=-0.001, \gamma_b=-0.0011$. The $N$ daughter frequencies, damping rates, and coupling coefficients are random values within one percent of $1.0$, $0.01$, and $1.0$, respectively.
\label{fig:MD}}
\end{figure*}

In this section, we restrict ourselves to networks that consist of only directly coupled modes. \citet{Dziembowski:82} carried out a stability analysis of directly coupled two-mode and three-mode systems. In the two-mode case, there is one self-coupled parent driving a single daughter.  In the three-mode case, there are two distinct parents driving a single daughter. \citet{Dziembowski:82} showed that the two-mode system can be stable if the magnitude of the parent driving rate is less than half that of the daughter damping rate.  However, he showed that the three-mode system is never stable.  In Appendix \ref{APP:A1} we carry out a stability analysis very similar to that of \citet{Dziembowski:82} and arrive at the same conclusion. We also extend the result to direct coupling involving two parents exciting $N\ge 1$ identical daughters (i.e., $N$ triplets; by identical we mean identical frequency and damping rate but not necessarily identical amplitude and phase). We find that the system is unstable even if $N\gg 1$,  suggesting that the excitation of many damped, directly-coupled daughters does not stabilize the system.

We carried out a series of numerical experiments that verify the results of the analytic stability analysis (and extends them by considering $N>1$ non-identical daughters, which is not amenable to analytic calculation).  
In the top panel of Figure~\ref{fig:AE_Simple_DC_double}, we consider a two-mode system consisting of a self-coupled parent mode $a=b$ driving a perfectly resonant daughter mode $c$ with a coupling coefficient $\kappa_{aac}=1.0$.  The parent frequency and driving rate are  $\omega_a=-0.5$ and $\gamma_a = -0.001$ , while the daughter frequency and damping rate are $\omega_c=1.0$ and $\gamma_c = 0.01$ (in this section, we use dimensionless time units).  Consistent with the stability analysis for a two-mode system, we find that the solution is stable and the parent settles into a nonlinear equilibrium at an energy (see Appendix \ref{APP:A1}) 
\begin{equation}
    E_{a, \rm eq} = \dfrac{\gamma_b\gamma_c}{4\kappa_{abc}^2\omega_b\omega_c}\left[1+\dfrac{\Delta^2}{\gamma^2}\right],
\label{eq:Ea_eq_DC}
\end{equation}
and similarly for the daughter, where $\gamma=\gamma_a+\gamma_b+\gamma_c$, $\Delta=\omega_a+\omega_b+\omega_c$, and here $a=b$ since the parent is self-coupled. 

In the bottom panel of Figure~\ref{fig:AE_Simple_DC_double}, we consider a three-mode system consisting of two distinct parent modes $a$ and $b$ and a daughter mode $c$.  The only difference between this three-mode system and the two-mode system shown in the top panel is that the parents have slightly different frequencies and damping rates ($\omega_a=-0.500$, $\omega_b=-0.502$; $\gamma_a=-0.0011$, $\gamma_b=-0.001$); the values of $\kappa_{abc}$ and the daughter parameters are otherwise the same.  Consistent with the stability analysis, we find that the system is now unstable, with one parent undergoing running away growth\footnote{The parent with the larger frequency or otherwise larger driving rate  is the one that tends to runaway.} (growing at its linear driving rate) while the other parent and daughter decay (they decay at about half the daughter's linear damping rate).  

In Figure~\ref{fig:MD} we consider two distinct parents directly coupled to $N>1$ identical daughters.  We consider $N=2$ in the left panel, $N=10$ in the middle panel, and $N=50$ in the right panel; the $N$ daughters have frequencies, damping rates, and $\kappa_{abc}$ that are random, uniformly distributed values within one percent of 1.0, 0.01, and 1.0, respectively, and thus are similar to the daughter parameters in Figure~\ref{fig:AE_Simple_DC_double}. We see that even with many linearly damped daughters weighing on the parents, the system remains unstable and that one parent always undergoes runaway growth while the other parent and the $N$ daughters decay (the rate of runaway growth and decay is nearly independent of $N$). This is consistent with the stability analysis given in Appendix~\ref{APP:A1} (which assumes identical daughters) and suggests that three-mode interactions involving only direct coupling (two distinct linearly driven parents coupled to $N\ge 1$ linearly damped daughters) is always unstable. We also see that as $N$ increases, the transition from growth to decay of the daughters and parent $a$ occurs earlier and thus they reach a smaller maximum energy.

\section{\bf M\lowercase{ixed} 
C\lowercase{oupling} E\lowercase{xperiments}}
\label{sec:direct + parametric}

In this section, we consider parents that simultaneously excite daughters through direct and parametric coupling, which we will call mixed coupling. Many studies of nonlinear mode coupling in stars have considered the impact of parametric coupling by itself  on linearly driven oscillation modes (often referred to as the three-mode parametric instability; see, e.g., \citealt{Wu:01, Brink:2005, Weinberg:21}).  Far fewer have considered the impact of direct coupling and none, to our knowledge, have considered both at the same time.  However, since both couplings can become dynamically important at large parent energies and because direct coupling alone is inherently unstable, it is worth considering  both simultaneously and examining how they influence one another. For these mixed networks we will focus on numerical solutions since the analytic stability analysis of even a four mode network is forbidding.  We will see, however, that the stability analyses of three-mode systems described above can provide useful insights into the behavior of mixed networks. 

In Section~\ref{sec:parametric}, we  provide a brief review of parametric coupling.  In Section~\ref{sec:experiments}, we describe the results of our numerical experiments with small mixed coupling networks. Finally, in Section~\ref{sec:mu}, we consider the effect mixed coupling has on the direct coupling strength parameter $\mu$ that \citet{Breger:14} measured in KIC 8054146.

\subsection{Parametric coupling}
\label{sec:parametric}
In parametric coupling, a single linearly driven parent excites a pair of resonant, linearly damped daughters.  Unlike direct coupling, the parent must exceed a threshold  amplitude in order to excite the daughters (it is therefore often referred to as a parametric instability).  Upon exceeding the threshold, the daughters begin to grow and eventually the system reaches a nonlinear equilibrium in which the mode energies are either constant (steady state) or undergo a chaotic limit cycle   (see, e.g., \citealt{Wersinger:1980, Dziembowski:82, Dimant:2000}); thus, unlike direct coupling, the three-mode system is stable.  Provided that the daughter damping rate is much larger than the magnitude of the parent driving rate, the equilibrium is a steady state if the magnitude of the frequency detuning $\Delta$ is larger than that of the daughter damping rate; otherwise the equilibrium is a limit cycle.  As shown in \citet{Dziembowski:82}, the expression for the  equilibrium energy of parametric coupling in steady state is the same as that of direct coupling and thus also given by Equation~(\ref{eq:Ea_eq_DC}).

\begin{figure}[t!]
\centering
\includegraphics[width=3.3in]{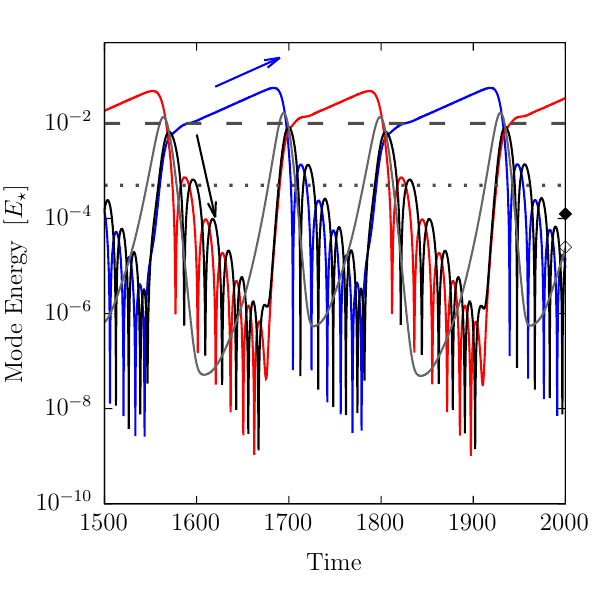}
\caption{Equilibrium solution of a mixed coupling network consisting of two distinct parents (red and blue lines), a directly coupled daughter (black line), and a self-coupled parametric daughter (grey line; i.e., one direct coupling triplet and two parametric coupling triplets). Listed respectively, their frequencies are $\{-1.000, -1.001, 2.0, 0.5\}$ and their linear driving or damping rates are $\{-0.01, -0.01, 0.1, 0.1\}$.  The coupling coefficient equals one for all three triplets. The horizontal lines show the parametric threshold energy (dashed line) and the equilibrium energy of the parametric daughter (dotted line). The filled mark on the right side of the plot indicates the  equilibrium energy of the direct daughter while the open mark indicates the equilibrium energy of the parents under direct coupling. The slope of the arrows indicate the parent driving rates (blue arrow) and the daughter damping rates (black arrow).
\label{fig:DC_GR}}
\end{figure}

\subsection{Experiments with small mixed coupling networks}
\label{sec:experiments}

We begin by considering small mixed coupling networks.  In Figure~\ref{fig:DC_GR} we show the energy as a function of time for a system that consists of four modes: two distinct parent modes $a$ and $b$ (blue and red lines), a direct daughter mode $c$ (black lines), and a self-coupled parametric daughter mode $d$ (grey lines). By Equation~(\ref{eq:amp eqn}),  the amplitude equations for this network are
\bea
    \dot{q}_a +(i\omega_a + \gamma_a)q_a &=&  i \omega_a (2 \kappa_{abc} q_b^* q_c^* + \kappa_{add} q_d^* q_d^*),\nonumber \\
    \dot{q}_b + (i\omega_b + \gamma_b) q_b &=& i\omega_b (2 \kappa_{abc} q_a^* q_c^* + \kappa_{bdd} q_d^* q_d^*), \nonumber \\
    \dot{q}_c + (i\omega_c +  \gamma_c) q_c &=& 2i \omega_c \kappa_{abc} q_a^* q_b^*, \nonumber \\
    \dot{q}_d + (i\omega_d + \gamma_d) q_d &=& 2i \omega_d (\kappa_{add} q_a^* q_d^* + \kappa_{bdd} q_b^* q_d^*).
\eea
In order $a$ through $d$, the mode frequencies are $\{1.000, 1.001, 2.0, 0.5\}$ and the driving and damping rates are $\{-0.01, -0.01, 0.1, 0.1\}$. The four coupling coefficients are all set equal to one.  The parameter values are chosen to be similar to those of Figure~\ref{fig:AE_Simple_DC_double} for comparison purposes.  We consider this network structure first because it is the smallest interesting network that includes both direct and parametric coupling.  It is interesting because in the absence of parametric coupling it would be unstable; although we could create a three-mode mixed network consisting of a self-coupled parent, a direct daughter, and a self-coupled parametric daughter, as we already saw such a network would be stable even in the absence of the parametric daughter. It therefore would not test whether parametric coupling can act to stabilize direct coupling. 

\begin{figure*}[t!]
\centering
\includegraphics[width=\textwidth]{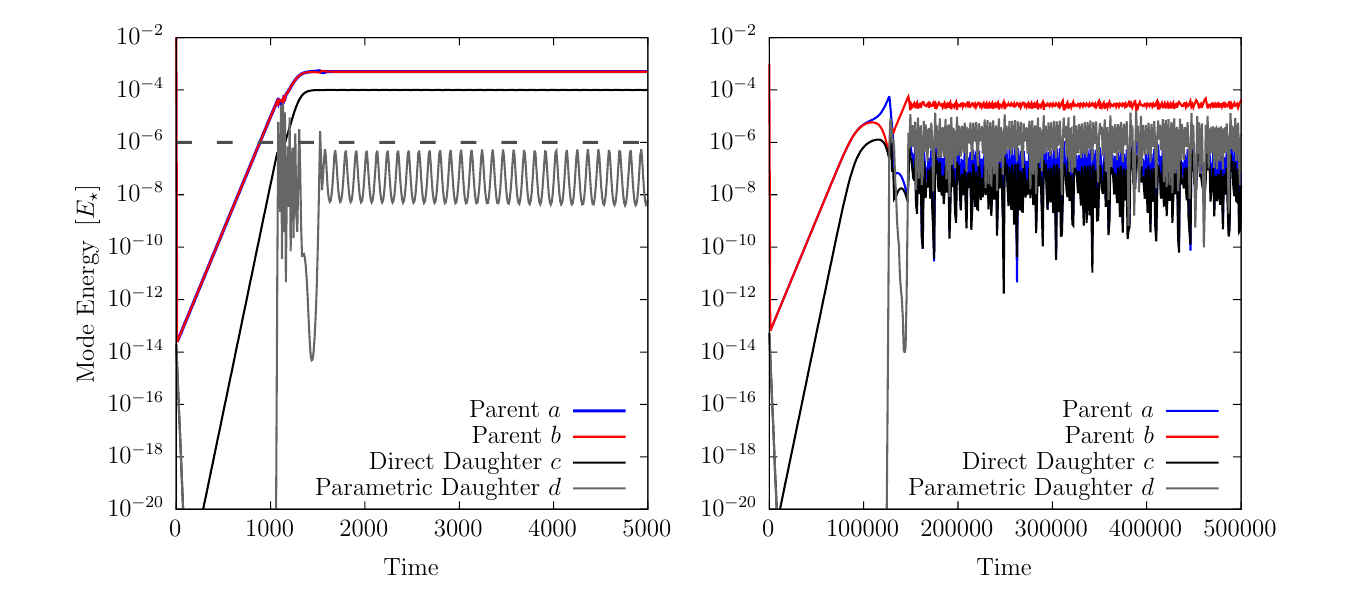}
\caption{Examples of different types of limit cycle behavior for four-mode mixed coupling networks. The network parameters are the same as in Figure~\ref{fig:DC_GR} except for the following differences. Left panel: the coupling coefficient $\kappa_{abc}$ for parametric coupling is 100 times larger than that for direct coupling (rather than being equal). Right panel:  the linear driving and damping rates are all 100 times smaller in magnitude. In the left panel, the two parents nearly overlap and the horizontal dashed line shows their parametric energy threshold. 
\label{fig:AE_SPS_double.pdf}}
\end{figure*}

The most notable feature of Figure~\ref{fig:DC_GR} is that the presence of the parametric daughter  stabilizes the system and prevents the runaway growth that occurs when only direct coupling is accounted for (compare Figure~\ref{fig:DC_GR} to the bottom panel of Figure~\ref{fig:AE_Simple_DC_double}). The evolution follows a relatively complex pattern that is very similar to the limit cycle behavior discussed in the context of parametric coupling only (see, e.g., \citealt{Wu:01}); for the most part, the directly coupled daughter is just pulled along for the ride.  

Specifically, we find that when a parent passes the parametric threshold (horizontal dashed line), the parametric daughter begins to quickly grow.  The latter eventually reaches such a large amplitude (for the chosen parameters, this occurs near the parent equilibrium energy of $\sim 0.01 E_\star$) that the parent's growth stops and it begins to decay rapidly.  Soon after, the parametric daughter reaches its maximum energy and once the parent drops below the parametric threshold the daughter begins to decay.  At nearly the same time, the other parent begins to rise  and the cycle repeats but with the other parent acting as the trigger that again destabilizes the parametric daughter (see the alternating red and blue lines at large energies). Meanwhile, the directly coupled daughter, following the cycle of the parents that drive it, also undergoes episodes of growth and decay.  The slope of the blue arrow shows that during the stage when a parent is above the parametric threshold, it grows at nearly the linear rate. The slope of the black arrow shows that the daughters decay at roughly the linear damping rate during the downward stage of the limit cycle (the decay is not exactly at the linear rate due to the complex nonlinear dynamics).  Finally, we see that the equilibrium energy of the direct coupling daughter (filled mark on right side of plot) provides a rough estimate of the characteristic energy about which the daughter oscillates.

In all of our numerical experiments, the four mode networks with mixed coupling exhibit cyclic behavior similar to that seen in Figure~\ref{fig:DC_GR}, although with patterns that depend on the parameter values.  In Figure~\ref{fig:AE_SPS_double.pdf} we show two other examples of the types of behavior we observe in networks that again consist of two parents, a direct daughter, and a self-coupled parametric daughter.

In the left panel, the key difference relative to Figure~\ref{fig:DC_GR} is that the coupling coefficient for parametric coupling is $100$ times larger than that for direct coupling (rather than being equal).  We see that the parents and direct daughter achieve a near steady state equilibrium that is well above the  the energy of the parametric daughter (which grows rapidly once the parents cross the parametric threshold, after which it settles into a regular pattern of oscillations). The average energy of the parametric daughter is significantly smaller than in Figure~\ref{fig:DC_GR} because the coupling coefficient is much larger (according to Equation~(\ref{eq:Ea_eq_DC}), the equilibrium energy scales as $\kappa_{abc}^{-2}$).
In the right panel of Figure~\ref{fig:AE_SPS_double.pdf}, the key difference relative to Figure~\ref{fig:DC_GR} is that all the linear driving and damping rates are a hundred times smaller in magnitude.  

The energies of the direct and parametric daughters are again well-separated except that now the former lies well below the latter. 

These results suggest that the  variability and energy of modes in mixed networks can be sensitive to the values of the linear and nonlinear mode parameters.

\begin{figure*}
\centering
\includegraphics[width=\textwidth]{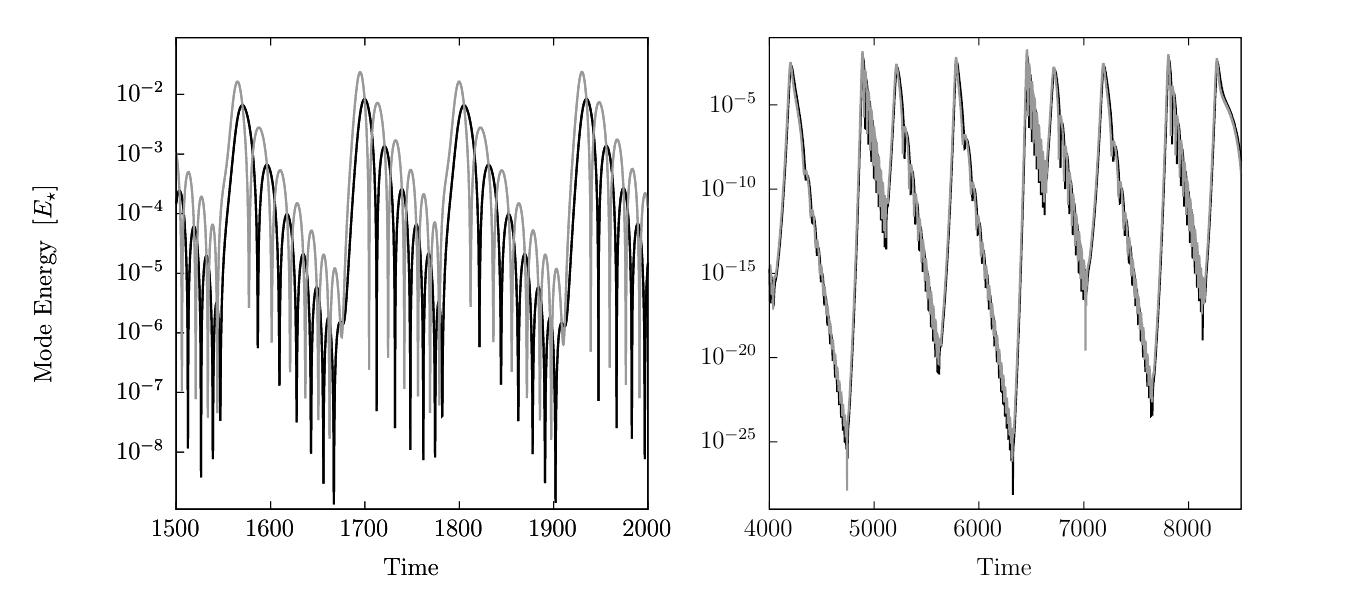}
\caption{
Energy of the direct daughter $E_c=|q_c|^2$ (black line) for a four-mode mixed coupling network  and its approximation $\mu^2 E_a E_b$ (grey line) as given by Equation~(\ref{eq:mu_amp_relation}), where $E_a$ and $E_b$ are the energies of the parents.The left panel uses the same mode parameter values as
Figure~\ref{fig:DC_GR} while the right panel uses parent driving rates that are 70\% smaller.
\label{fig:2D}}
\end{figure*}

\subsection{Measuring the coupling strength parameter $\mu$ in networks with mixed coupling}
\label{sec:mu}

The oscillation spectra of the \dsct{} star KIC 8054146 contains daughter modes whose amplitudes $q_c$ appear to  vary in time as 
\beq
|q_c(t)| = \mu |q_a(t)| |q_b(t)|,
\label{eq:mu_amp_relation}
\eeq
where $q_a$ and $q_b$ are the amplitudes of the parent modes and $\mu$ is the constant coupling strength parameter given by
\beq
\mu = \dfrac{\left|\omega_c \kappa_{abc}\right|}{\sqrt{\Delta^2 + \gamma_c^2}}.
\label{eq:mu}
\eeq
\citet{Breger:14} use the observed temporal correlation between daughter and parent amplitudes  to directly measure $\mu$ for a number of triplets in KIC 8054146 and find values as large as $\mu\sim 10^4$. 

In \citetalias{Mourabit:23}, we analyzed the amplitude equations for three-mode direct coupling and showed that Equation~(\ref{eq:mu_amp_relation}) should be a good approximation for $|q_c(t)|$ as long the daughter energy is less than that of the parents.  However, our analysis did not consider the potential impact of parametric coupling on the amplitude evolution of the modes.  

As we illustrate in Figure~\ref{fig:2D},  the relation should nonetheless still be a good approximation  even in the presence of mixed coupling.  The black lines show the numerically integrated energy $E_c(t)=|q_c|^2$ of a directly coupled daughter while the orange lines show $\mu^2 E_a E_b$. The left panel uses the same mode parameter values as in Figure~\ref{fig:DC_GR}, while the right panel uses parent driving rates that are  only thirty percent of the values in Figure~\ref{fig:DC_GR}.  We see that the time-variation of the two lines are quite similar, especially for the smaller parent driving rates in the right panel.  This suggests that Equation~(\ref{eq:mu_amp_relation}) should approximately hold in the presence of mixed coupling  and that Equation~(\ref{eq:mu}) provides a good estimate of $\mu$, although how accurate it is may depend on the particular values of the linear and nonlinear parameters. How useful these relations are in larger mixed networks as may be found in \dsct{} stars remains to be seen, however.

\section{\bf M\lowercase{ixed} C\lowercase{oupling in} \dsct{} S\lowercase{tars}}
\label{sec:delta scuti}

In the previous section, we considered simple mixed coupling networks with mode parameters chosen by hand in order to explore how the network dynamics and stability depends on the parameter values. In this section, we again consider simple mixed coupling networks but we now use mode parameters calculated from a realistic model of a \dsct{} star (based on the calculations of \citetalias{Mourabit:23}, as described below). The primary purpose is to demonstrate that mixed coupling can occur within \dsct{} stars and impact the time-variation of modes driven by   the $\kappa$ mechanism. Since a realistic star can support a very large number of linearly and nonlinearly excited modes, the results shown below, while informative, do not capture the complicated  multi-mode nonlinear interactions that can occur.  In future work, we plan to build on this study and construct more realistic mixed coupling networks whose dynamics and oscillation spectra can then be compared with observed \dsct{} stars.

In \citetalias{Mourabit:23}, we used the stellar evolution code \texttt{MESA} \citep{Paxton:11, Paxton:13, Paxton:15, Paxton:18, Paxton:19, Jermyn:23} to construct 14 \dsct{} models that span the instability strip. We then used the stellar oscillation code \texttt{GYRE} \citep{Townsend:13, Townsend:18} to find linear eigenmodes for each model with frequencies and angular degrees within the range observed by \emph{Kepler}. By searching among these modes for directly coupled triplets with large $\mu$, we found that $\mu$ values as large as \citet{Breger:14} detected in KIC 8054146 were common among the 14 models. 

\begin{figure*}[t!]
\centering
\includegraphics[width=\textwidth]{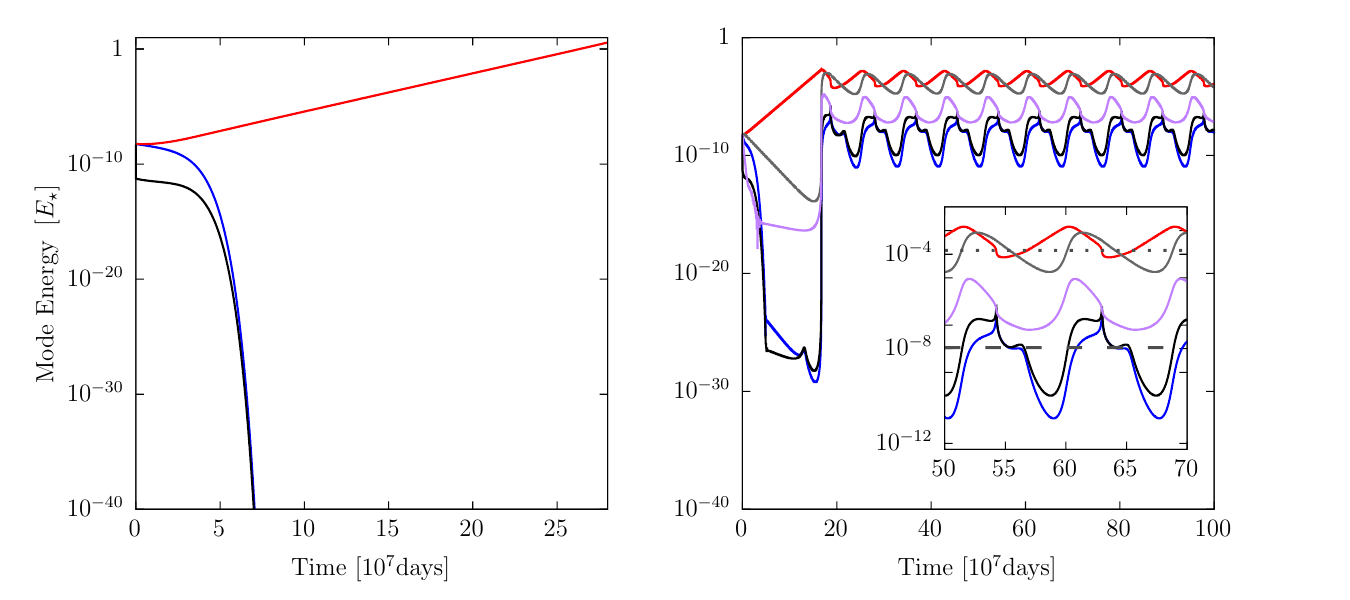}
\caption{Mode energy as a function of time for linear and nonlinear mode parameters calculated from a model of a \dsct{} star.  The parameter values are given in the main text; the parents are red and blue, the direct daughter is black, and the parametric daughters are grey and purple. The left panel shows the solution for a three mode network that consists only of a direct coupling triplet, and as a result is unstable. The right panel shows the solution for a five mode mixed coupling network.  It includes the same direct daughter as the left panel but now also allows the parents to couple to parametric daughters, and as a result the system is stabilized and settles into a limit cycle equilibrium. The inset plot zooms in on a few cycles and shows the parametric threshold for the red parent (dotted line) and the blue parent (dashed line). 
\label{fig:Dsct2_0_double}}
\end{figure*}

In Table 1 of \citetalias{Mourabit:23}, we list  sets of triplets with large values of $\mu$ for each \dsct{} model.  For the calculations presented in this section, we choose\footnote{While this choice of triplet is essentially arbitrary, it offers practical advantages over some of the others listed in the table.  First, the parents are linearly driven according to \texttt{GYRE}'s solutions of the non-adiabatic linear equations, which is not true of many of the large $\mu$ triplets listed in the table (as explained in \citetalias{Mourabit:23}, it is not clear why only a small subset of modes found with \texttt{GYRE} are linearly unstable while such modes appear relatively common in the observed spectra of \dsct{} stars). Second, the parents are distinct modes that have similar frequencies.  Distinct parents, as opposed to self-coupled parents, allow us to test whether parametric coupling stabilizes direct coupling for realistic mode parameters.  The advantage of choosing parents with similar frequencies is that the parents can then have relatively small detunings $\Delta$ with the same parametric daughter pair (the angular selection rules also then require that the angular degree of the parents have the same parity); this ensures that the parametric threshold energy is small for both triplets without requiring two separate daughter pairs for each parent (which would expand the network to seven modes rather than five).  We want parents with small parametric threshold energies because when they cross it and the daughters begin to grow,
the nonlinear growth rates, and thus the numerical integration times, remain reasonably small (since the growth rate is  proportional to parent amplitude).} the direct coupling triplet listed fifth for the model with $M = 2.0 M_\sun$, $T_{\mathrm{eff}} = 7202 \textrm{ K}$, and $\log g = 3.80$.  The parameters of the two parents and direct daughter are, respectively, $l=\{3,1,2\}$, $n=\{0,3,8\}$, $\omega=\{1.17, 1.27, 2.42\}\times 10^{-3} \textrm{ rad s}^{-1}$, $\gamma=\{-4.8\times10^{-9}, -1.1\times10^{-8}, 2.8\times10^{-5}\}\textrm{ s}^{-1}$. Their coupling coefficient is $\kappa_{abc}=6.3$ and their detuning is $\Delta=7.6\times10^{-5}\textrm{ rad s}^{-1}$; by Equation~(\ref{eq:mu}), these parameters give $\mu\approx 2000$

Given this direct triplet, we search for a single parametric daughter pair to couple to both parents. The search is carried out with \texttt{GYRE} over all modes with $l < 25$ and $|n| < 400$ that satisfy the angular selection rules for three-mode coupling and have small $\Delta$.  Note that although \emph{Kepler} cannot resolve modes with $l \gtrsim 3$, our search includes higher $l$ modes because they can nonetheless couple to the low $l$ parents and influence their dynamics. The parameters of the parametric daughter pair we identify and use in our mixed coupling network are: $l=\{14,15\}$, $n=\{-282, -12\}$, $\omega=\{1.09, 0.0822\}\times10^{-3}\textrm{ rad s}^{-1}$, and $\gamma=\{1.4, 1.5\}\times 10^{-5}\textrm{ s}^{-1}$.  The pair couples to each parent with coupling coefficient $\kappa=\{0.85, 3.85\}$ and detuning $\Delta=\{9.8\times10^{-8}, 1.0\times10^{-4}\} \textrm{ rad s}^{-1}$.

The left panel of Figure~\ref{fig:Dsct2_0_double} shows mode energy as a function of time for the direct coupling triplet only (the two parents and the directly coupled daughter). As expected (see Section~\ref{sec:stability}), the system is unstable, with one of the parents growing without bound at nearly its linear driving rate.  The right panel shows the results for the full five-mode mixed coupling network that includes the direct daughter and both parametric daughter pairs.  The system is now stable and executes limit cycle oscillations similar to those found in the hand-crafted networks of Section~\ref{sec:stability}.  The inset zooms in on the cyclic pattern and shows the parametric threshold energies for the two parents. We find that the mode amplitudes vary on timescales of $\gtrsim 10^6\textrm{ days}$, which is $\sim 100$ times longer than the fastest amplitude variations observed in \dsct{} stars \citep{Breger:14, Bowman:16}.  However, this could be because of the limited size of our network.  Lastly, although the limit cycle pattern is fairly complicated, we see that the parents roughly oscillate about their respective threshold energies. The latter thus provide a rough estimate of the characteristic parent  energy.  It will be interesting to see if this remains true in future studies that include larger, more realistic mode networks, as it could help simplify the interpretation of observed spectra that include nonlinearly interacting modes.

\section{\bf S\lowercase{ummary} \lowercase{and} C\lowercase{onclusions}}
\label{sec:conclusions}
Motivated by the evidence of three-mode coupling observed in the oscillation spectra of many \dsct{} stars, and especially the detection of directly coupled modes in KIC 8054146, we studied the stability and dynamics of modes interacting jointly via direct and parametric coupling.  We showed that direct coupling by itself is inherently unstable, with one parent always diverging in amplitude even if the parents excite an arbitrarily large number of  directly coupled daughters. We found, however, that if these same parents also excite daughters through parametric coupling, there exist stable solutions even for relatively small mode networks (e.g., two parents, a direct daughter, and two parametric daughters).  By integrating small mode networks using linear and nonlinear mode parameters calculated from a realistic \dsct{} model, we found that direct and parametric coupling are likely to be simultaneously active within \dsct{} stars. Notably, the presence of parametric coupling does not necessarily disrupt the  correlations in amplitude and phase between directly coupled modes, such as those \citet{Breger:14} found in their analysis of KIC 8054146.

We limited our study to small mode networks consisting of two parents and two or three daughters (with the exception of the direct-only analysis in Section~\ref{sec:experiments} which considered up to 50 daughters).  This allowed us to explore the minimum conditions needed to stabilize mixed coupling networks (those subject to both direct and parametric coupling) and to study how different linear and nonlinear mode parameters impact the mode dynamics. In an actual \dsct{} star, there can be hundreds of excited modes with detectable amplitudes (see, e.g., \citealt{Balona:2015}).  Thus, future studies of mode coupling will need to consider much larger mode networks in order to compare with observations of \dsct{} stars. This is likely to be a challenging problem.  It requires constructing and integrating large sets of coupled modes,  ensuring a large enough network is used by testing for convergence, and accounting for the relation between the intrinsic amplitudes of  modes (i.e., $q_a$) and their observed flux variations. Although the latter relation is known in principle \citep{Dziembowski:77, Watson:1988}, the results can be sensitive to how the regions near the photosphere are treated \citep{Pfahl:08}. 

When looking for evidence of nonlinear mode coupling in the oscillation spectra of \dsct{} stars, some studies have attempted to count the number of resonant triplet combinations found among the observed modes (see, e.g., \citealt{Balona:2024}). However, it is worth noting that an observed mode can be nonlinearly coupled to other modes even if it does not form a resonant triplet with other \emph{observed} modes. This can happen for two reasons.  First, the observed mode can be coupled to a pair of modes that are below the detection threshold; for example, if the triplet is undergoing limit cycle oscillations and only one mode of the triplet happens to have large amplitude at any given time (an example of this can be seen in Figure~\ref{fig:DC_GR}).  Second, the three-mode angular selection rules allow low-degree modes ($l \lesssim 3$, say) to couple to high-degree modes ($l \gg 3$).  Since telescopes such as \emph{Kepler} and \emph{TESS} can only resolve modes with $l\lesssim 3$, they would not detect the high-degree modes that might be coupled to the observed low-degree mode. 

The only indication that a mode is nonlinearly interacting with other modes might therefore be through modulations in its amplitude, phase, or frequency.\\

This work was supported by NASA ATP grant 80NSSC21K0493.

\software{\texttt{MESA} (\citealt[][\url{http://mesa.sourceforge.net}]{Paxton:11, Paxton:13, Paxton:15, Paxton:18,Paxton:19, Jermyn:23}),
\texttt{GYRE} (\citealt[][\url{https://gyre.readthedocs.io/en/stable/}]{Townsend:13, Townsend:18}).}

\appendix

\section{\bf D\lowercase{irect} C\lowercase{oupling }S\lowercase{tability} A\lowercase{nalysis}}
\label{APP:A1}

In this appendix, we analyze the stability of directly coupled systems in which there are two distinct linearly driven parents coupled to $N\ge 1$ identical daughters. We first consider the $N=1$ case and show later that the stability of $N>1$ daughters follows simply from the $N=1$ result. We begin by finding the equilibrium solution, which we then perturb to evaluate its stability.  Our analysis is very similar to  that given in \citet{Dziembowski:82}, the main difference being that we use our notation and normalization, show some of the steps in greater detail, and extend the result to $N>1$.

Let modes $a$ and $b$ be linearly driven parents ($\gamma_{a,b} < 0$) and let mode $c$ be a linear damped daughter ($\gamma_c > 0$). By Equation~(\ref{eq:amp eqn}), the amplitude equations for the system are
\beq
    \dot{q}_a + (i\omega_a + \gamma_a)q_a = i \omega_a \sum_{b}\sum_{c} \kappa_{abc} q^*_b q^*_c
     = i\omega_a \kappa q^*_b q^*_c,
\eeq
where $\kappa = 2 \kappa_{abc}$ and here and below we do not show the expressions for the other two modes since they are identical up to permutations of the mode indices. For each mode $j\in \{a,b,c\}$, apply a change of coordinates $q_{j}(t) = X_j(t) e^{-i\omega_j t}$ and use amplitude-phase form $X_j(t)=  \varepsilon_j(t) e^{i \alpha_j(t)}$ with $\varepsilon_j, \alpha_j \in\mathbb{R}$, to get
\beq
    \dot{\varepsilon}_a + (i \dot{\alpha}_a + \gamma_a) \varepsilon_a = i \omega_a \kappa \varepsilon_b \varepsilon_c e^{-i\beta}, 
\eeq
where $\beta(t) = \alpha_a + \alpha_b + \alpha_c - t\Delta $ and $\Delta = \omega_a + \omega_b + \omega_c$ is the frequency detuning. Separating the real and imaginary parts,
\beq
 \dot{\varepsilon}_a + \gamma_a \varepsilon_a = \omega_a \kappa \varepsilon_b \varepsilon_c \sin\beta,
 \hspace{1.0cm}
 \dot{\alpha}_a\varepsilon_a  = \omega_a \kappa \varepsilon_b \varepsilon_c \cos\beta,
 \label{eq:doteps}
\eeq

dividing these equations by $\varepsilon_a$ (and similarly by $\varepsilon_b$ and $\varepsilon_c$ for the version of this equation for the other two modes), and adding the results from all three modes gives
\bea
    \dfrac{\mathrm{d} \ln \left(\varepsilon_a\varepsilon_b\varepsilon_c\right)}{\mathrm{d}t} + \gamma &=& \left ( \omega_a \dfrac{\varepsilon_b \varepsilon_c}{\varepsilon_a} + \omega_b \dfrac{\varepsilon_a \varepsilon_c}{\varepsilon_b} + \omega_c \dfrac{\varepsilon_a \varepsilon_b}{\varepsilon_c}  \right ) \kappa \sin \beta, \label{eq:lneps}\\
    \dot{\beta}+\Delta &=&  \left ( \omega_a \dfrac{\varepsilon_b \varepsilon_c}{\varepsilon_a} + \omega_b \dfrac{\varepsilon_a \varepsilon_c}{\varepsilon_b} + \omega_c \dfrac{\varepsilon_a \varepsilon_b}{\varepsilon_c}  \right )\kappa \cos \beta,
    \label{eq:betadot plus Delta}
\eea
where $\gamma = \gamma_a + \gamma_b + \gamma_c$. Dividing Equation~(\ref{eq:betadot plus Delta}) by Equation~(\ref{eq:lneps}), we get
\beq
\dot{\beta} =  \cot \beta\left [ \dfrac{\mathrm{d} \ln \left(\varepsilon_a\varepsilon_b\varepsilon_c\right)}{\mathrm{d}t} + \gamma \right ] - \Delta.
\label{eq:betadot}
\eeq
This equation together with the equation on the left of line (\ref{eq:doteps}) and the corresponding ones for modes $b$ and $c$ are equivalent to Equations (6.3)-(6.5) of \citet{Dziembowski:82}. 

The equilibrium solution is found by setting the time derivatives to zero (since the equilibrium corresponds to constant mode amplitude).  Equations~(\ref{eq:doteps}) and (\ref{eq:betadot}) then give 
\beq
\varepsilon^2_{a, \mathrm{eq}} =  \dfrac{\gamma_b \gamma_c}{4\omega_b \omega_c \kappa_{abc}^2} \left[1 + \dfrac{\Delta^2}{\gamma^2}\right], \label{EQ:DCEQ1} 
\hspace{0.5cm}
\cot \beta_{\mathrm{eq}} = \dfrac{\Delta}{\gamma},
\eeq
and similarly for modes $b$ and $c$, where $\varepsilon^2_{j, \mathrm{eq}}$ are the equilibrium energies, which in Equation~(\ref{eq:Ea_eq_DC}) we write as $E_{a, \rm eq}$.

In order to analyze the stability of the equilibrium solution, introduce infinitesimal perturbations to the amplitudes $\varepsilon_j = \varepsilon_{j, \rm eq} + \delta \varepsilon_j$ and phase $\beta = \beta_{\rm eq} + \delta \beta$, and  substitute them into Equations~(\ref{eq:doteps}). Keeping only terms that are linear in the small perturbations gives 
\bea
    \dfrac{\mathrm{d}}{\mathrm{d}t} \left ( \dfrac{\delta \varepsilon_a}{\varepsilon_{a, \mathrm{eq}}} \right) &=& \gamma_a \left [ \eta \delta \beta + \dfrac{\delta \varepsilon_b}{\varepsilon_{b, \mathrm{eq}}} + \dfrac{\delta \varepsilon_c}{\varepsilon_{c, \mathrm{eq}}} - \dfrac{\delta \varepsilon_a}{\varepsilon_{a, \mathrm{eq}}} \right] \label{eq:ddt_deps}\\
      \dfrac{\mathrm{d} \delta \beta}{\mathrm{d}t} &=& \eta \left [ 
    \dfrac{\mathrm{d}}{\mathrm{d} t} \left ( \dfrac{\delta \varepsilon_a}{\varepsilon_{a, \mathrm{eq}}} \right ) + \dfrac{\mathrm{d}}{\mathrm{d} t} \left (\dfrac{\delta \varepsilon_b}{\varepsilon_{b, \mathrm{eq}}} \right ) + \dfrac{\mathrm{d}}{\mathrm{d} t} \left ( \dfrac{\delta \varepsilon_c}{\varepsilon_{c, \mathrm{eq}}} \right )
    \right] - (1 + \eta^2) \gamma \delta \beta
\eea
If we now assume the perturbations have a time dependence $e^{st}$, the set of four perturbation equations can be written as
\begin{align}
   \begin{pmatrix}
        s + \gamma_a & -\gamma_a & -\gamma_a \\
        -\gamma_b & s + \gamma_b & -\gamma_b \\
        -\gamma_c & -\gamma_c & s + \gamma_c
    \end{pmatrix}
    \begin{bmatrix}
        \delta \varepsilon_a / \varepsilon_{a, \mathrm{eq}}\\
        \delta \varepsilon_b / \varepsilon_{b, \mathrm{eq}}\\
        \delta \varepsilon_c / \varepsilon_{c, \mathrm{eq}}
    \end{bmatrix}
    = \eta \delta \beta 
    \begin{bmatrix}
        \gamma_a\\
        \gamma_b\\
        \gamma_c\\
    \end{bmatrix}
\end{align}
and
\begin{align}
    s \left ( \dfrac{\delta \varepsilon_a}{\varepsilon_{a, \mathrm{eq}}} + \dfrac{\delta \varepsilon_b}{\varepsilon_{b, \mathrm{eq}}} + \dfrac{\delta \varepsilon_c}{\varepsilon_{c, \mathrm{eq}}}\right ) \eta - \delta \beta \big{[}\gamma(1 + \eta^2) + s\big{]} = 0,
    \label{eq:sdeps}
\end{align}
which agrees with Equations (6.11)-(6.12) in \citeauthor{Dziembowski:82} (1982; note that we use the opposite sign convention for the driving and damping rates).
In order to write the equation in the form of a standard eigenvalue problem $\textbf{H}\gv{R} = s \gv{R}$, we can add the three versions of Equation~(\ref{eq:ddt_deps}) for each mode and substitute the result into Equation~(\ref{eq:sdeps}).    This gives 
\begin{align}
       \begin{pmatrix}
        s + \gamma_a & -\gamma_a & -\gamma_a & - \eta \gamma_a\\
        -\gamma_b & s + \gamma_b & -\gamma_b & - \eta \gamma_b \\
        -\gamma_c & -\gamma_c & s + \gamma_c & - \eta \gamma_c\\
        \left(2\gamma_a - \gamma\right) \eta  & 
        \left(2\gamma_b - \gamma\right) \eta & 
        \left(2\gamma_c-\gamma\right) \eta &
        s + \gamma
    \end{pmatrix}
    \begin{bmatrix}
        \delta \varepsilon_a / \varepsilon_{a, \mathrm{eq}}\\
        \delta \varepsilon_b / \varepsilon_{b, \mathrm{eq}}\\
        \delta \varepsilon_c / \varepsilon_{c, \mathrm{eq}}\\
        \delta \beta
    \end{bmatrix}
    = 0,
    \label{eq:eigen}
\end{align}
where the matrix corresponds to $s \textbf{I} -\textbf{H}$. By solving $\det\left( \textbf{H} - s \textbf{I} \right) = 0$, we derive the characteristic equation of the system
\beq
    s^4 + a_1 s^3 + a_2 s^2 + a_3 s + a_4 = 0,
\eeq
where the coefficients of the polynomial are
\bea
    a_1 &=& 2\gamma,
\hspace{3.5cm}
    a_2 = \gamma^2 (1 + \eta^2) - 4 (\gamma_a \gamma_b + \gamma_b \gamma_c + \gamma_a \gamma_c ) \eta^2, \nonumber \\
    a_3 &=& -4 \gamma_a\gamma_b\gamma_c (1 + 3 \eta^2),
    \hspace{1.0cm}
    a_4 = -4 \gamma\gamma_a\gamma_b\gamma_c (1+\eta^2).
\eea
The system is stable if it satisfies the Hurwitz criteria
\bea
    W_1 &=& a_1 > 0, \hspace{78pt} W_2 = a_1 a_2 - a_3 > 0, \nonumber \\
    W_3 &=& a_3 W_2 - a_1^2 a_4 > 0, \hspace{32pt} W_4 = a_4 W_3 > 0.
\eea

For a directly coupled system of three-distinct modes, the Hurwitz criteria cannot be satisfied.  To see this, suppose $W_4 > 0$. This implies $a_4>0$ if we are to also satsify the stability condition $W_3>0$. However, since the parent modes are linearly driven with $\gamma_a, \gamma_b < 0$ and the daughter mode is linearly damped with $\gamma_c > 0$, then by the expression for $a_4$ given above we see that $a_4>0$ requires $\gamma = \gamma_a + \gamma_b + \gamma_c < 0$. But then this implies $a_1<0$ and we cannot satisfy the stability requirement $W_1>0$.  A directly coupled three-mode system must therefore be unstable.

This result can be generalized to $N$ directly coupled identical daughters through a simple transformation. Let the two parent modes $a$ and $b$ couple to $N$ identical daughters with amplitudes $q_{c,1},...,q_{c,N}$. Since the daughters are assumed to be identical, their frequencies $\omega_c$, linear damping rates $\gamma_c$, and coupling coefficients $\kappa_{abc}$ are all the same (their initial conditions and hence amplitudes are not assumed to be identical, however). We can therefore write the parent amplitude equations as
\bea
    \dot{q}_a + (i \omega_a + \gamma_a)q_a &=& i \omega_a  \kappa_{abc} q_b^* \sum_i q_{c, i}^*,\\
    \dot{q}_b + (i \omega_b + \gamma_b)q_b &=& i \omega_b  \kappa_{abc} q_a^* \sum_i q_{c, i}^* 
\eea
and the sum of the daughter amplitude equations as
\beq
    \sum_i \dot{q}_{c, i} + (i \omega_c + \gamma_c)\sum_i q_{c, i} = i \omega_c  N \kappa_{abc} q_a^* q_b^*,
\eeq
where the sums run over all $N$ daughter modes.
If we now let $r_a=\sqrt{N} q_a$, $r_b=\sqrt{N}q_b$, and $r_c = \sum\limits_{i} q_{c, i}$ we get
\bea
    \dot{r}_a + (i \omega_a + \gamma_a) r_a &=& i \omega_a \kappa_{abc} r_b^* r_c^*, \\
    \dot{r}_b + (i \omega_b + \gamma_b) r_b &=& i \omega_b \kappa_{abc} r_a^* r_c^*, \\
    \dot{r}_c + (i \omega_c + \gamma_c) r_c &=& i \omega_c \kappa_{abc} r_a^* r_b^*.
\eea
Since this set of three equations has the exact same form as the original system of three directly coupled modes considered above, the equilibrium solution of these equations must likewise be unstable.

\bibliography{Paper2_v5}

\begin{thebibliography}{}
\expandafter\ifx\csname natexlab\endcsname\relax\def\natexlab#1{#1}\fi
\providecommand{\url}[1]{\href{#1}{#1}}
\providecommand{\dodoi}[1]{doi:~\href{http://doi.org/#1}{\nolinkurl{#1}}}
\providecommand{\doeprint}[1]{\href{http://ascl.net/#1}{\nolinkurl{http://ascl.net/#1}}}
\providecommand{\doarXiv}[1]{\href{https://arxiv.org/abs/#1}{\nolinkurl{https://arxiv.org/abs/#1}}}

\bibitem[{{Balona}(2024)}]{Balona:2024}
{Balona}, L.~A. 2024, The Open Journal of Astrophysics, 7, 5, \dodoi{10.21105/astro.2109.12574}

\bibitem[{{Balona} {et~al.}(2015){Balona}, {Daszy{\'n}ska-Daszkiewicz}, \& {Pamyatnykh}}]{Balona:2015}
{Balona}, L.~A., {Daszy{\'n}ska-Daszkiewicz}, J., \& {Pamyatnykh}, A.~A. 2015, Monthly Notices of the Royal Astronomical Society, 452, 3073, \dodoi{10.1093/mnras/stv1513}

\bibitem[{{Bowman} {et~al.}(2021){Bowman}, {Hermans}, {Daszy{\'n}ska-Daszkiewicz}, {Holdsworth}, {Tkachenko}, {Murphy}, {Smalley}, \& {Kurtz}}]{Bowman:21}
{Bowman}, D.~M., {Hermans}, J., {Daszy{\'n}ska-Daszkiewicz}, J., {et~al.} 2021, \mnras, 504, 4039, \dodoi{10.1093/mnras/stab1124}

\bibitem[{{Bowman} {et~al.}(2016){Bowman}, {Kurtz}, {Breger}, {Murphy}, \& {Holdsworth}}]{Bowman:16}
{Bowman}, D.~M., {Kurtz}, D.~W., {Breger}, M., {Murphy}, S.~J., \& {Holdsworth}, D.~L. 2016, \mnras, 460, 1970, \dodoi{10.1093/mnras/stw1153}

\bibitem[{{Breger}(1979)}]{Breger:1979}
{Breger}, M. 1979, Publications of the Astronomical Society of the Pacific, 91, 5, \dodoi{10.1086/130433}

\bibitem[{{Breger}(2000)}]{Breger:00}
{Breger}, M. 2000, in Astronomical Society of the Pacific Conference Series, Vol. 210, Delta Scuti and Related Stars, ed. M.~{Breger} \& M.~{Montgomery}, 3

\bibitem[{{Breger} \& {Montgomery}(2014)}]{Breger:14}
{Breger}, M., \& {Montgomery}, M.~H. 2014, \apj, 783, 89, \dodoi{10.1088/0004-637X/783/2/89}

\bibitem[{{Breger} \& {Pamyatnykh}(1998)}]{Breger:98}
{Breger}, M., \& {Pamyatnykh}, A.~A. 1998, Astronomy and Astrophysics, 332, 958.
\newblock \doarXiv{astro-ph/9802076}

\bibitem[{{Brink} {et~al.}(2004){Brink}, {Teukolsky}, \& {Wasserman}}]{Brink:2004}
{Brink}, J., {Teukolsky}, S.~A., \& {Wasserman}, I. 2004, \prd, 70, 121501, \dodoi{10.1103/PhysRevD.70.121501}

\bibitem[{{Brink} {et~al.}(2005){Brink}, {Teukolsky}, \& {Wasserman}}]{Brink:2005}
---. 2005, prd, 71, 064029, \dodoi{10.1103/PhysRevD.71.064029}

\bibitem[{{Buchler}(2009)}]{Buchler:2009}
{Buchler}, J.~R. 2009, in American Institute of Physics Conference Series, Vol. 1170, Stellar Pulsation: Challenges for Theory and Observation, ed. J.~A. {Guzik} \& P.~A. {Bradley} (AIP), 51--58, \dodoi{10.1063/1.3246556}

\bibitem[{{Buchler} {et~al.}(1997){Buchler}, {Goupil}, \& {Hansen}}]{Buchler:97}
{Buchler}, J.~R., {Goupil}, M.~J., \& {Hansen}, C.~J. 1997, Astronomy and Astrophysics, 321, 159

\bibitem[{{Chevalier}(1971)}]{Chevalier:1971}
{Chevalier}, C. 1971, Astronomy and Astrophysics, 14, 24

\bibitem[{{Cox}(1963)}]{Cox:1963}
{Cox}, J.~P. 1963, The Astrophysical Journal, 138, 487, \dodoi{10.1086/147661}

\bibitem[{{De Somma} {et~al.}(2020){De Somma}, {Marconi}, {Molinaro}, {Cignoni}, {Musella}, \& {Ripepi}}]{DeSomma:2020}
{De Somma}, G., {Marconi}, M., {Molinaro}, R., {et~al.} 2020, The Astrophysical Journal Supplement Series, 247, 30, \dodoi{10.3847/1538-4365/ab7204}

\bibitem[{{Dimant}(2000)}]{Dimant:2000}
{Dimant}, Y.~S. 2000, Physical Review Letters, 84, 622, \dodoi{10.1103/PhysRevLett.84.622}

\bibitem[{{Dziembowski}(1977)}]{Dziembowski:77}
{Dziembowski}, W. 1977, Acta Astronomica, 27, 203

\bibitem[{{Dziembowski}(1982)}]{Dziembowski:82}
---. 1982, Acta Astronomica, 32, 147

\bibitem[{{Dziembowski} \& {Krolikowska}(1985)}]{Dziembowski:85}
{Dziembowski}, W., \& {Krolikowska}, M. 1985, Acta Astronomica, 35, 5

\bibitem[{{Dziembowski} {et~al.}(1988){Dziembowski}, {Krolikowska}, \& {Kosovichev}}]{Dziembowski:88}
{Dziembowski}, W., {Krolikowska}, M., \& {Kosovichev}, A. 1988, Acta Astronomica, 38, 61

\bibitem[{{Essick} \& {Weinberg}(2016)}]{Essick:2016}
{Essick}, R., \& {Weinberg}, N.~N. 2016, The Astrophysical Journal, 816, 18, \dodoi{10.3847/0004-637X/816/1/18}

\bibitem[{{Gautschy} \& {Saio}(1995)}]{Gautschy:1995}
{Gautschy}, A., \& {Saio}, H. 1995, Annual Review of Astronomy and Astrophysics, 33, 75, \dodoi{10.1146/annurev.aa.33.090195.000451}

\bibitem[{{Guzik}(2021)}]{Guzik:21}
{Guzik}, J.~A. 2021, Frontiers in Astronomy and Space Sciences, 8, 55, \dodoi{10.3389/fspas.2021.653558}

\bibitem[{{Handler}(2009)}]{Handler:2009}
{Handler}, G. 2009, in American Institute of Physics Conference Series, Vol. 1170, Stellar Pulsation: Challenges for Theory and Observation, ed. J.~A. {Guzik} \& P.~A. {Bradley} (AIP), 403--409, \dodoi{10.1063/1.3246528}

\bibitem[{{Jermyn} {et~al.}(2023){Jermyn}, {Bauer}, {Schwab}, {Farmer}, {Ball}, {Bellinger}, {Dotter}, {Joyce}, {Marchant}, {Mombarg}, {Wolf}, {Sunny Wong}, {Cinquegrana}, {Farrell}, {Smolec}, {Thoul}, {Cantiello}, {Herwig}, {Toloza}, {Bildsten}, {Townsend}, \& {Timmes}}]{Jermyn:23}
{Jermyn}, A.~S., {Bauer}, E.~B., {Schwab}, J., {et~al.} 2023, \apjs, 265, 15, \dodoi{10.3847/1538-4365/acae8d}

\bibitem[{{Mourabit} \& {Weinberg}(2023)}]{Mourabit:23}
{Mourabit}, M., \& {Weinberg}, N.~N. 2023, \apj, 950, 6, \dodoi{10.3847/1538-4357/acca16}

\bibitem[{{Pamyatnykh}(1999)}]{Pamyatnykh:1999}
{Pamyatnykh}, A.~A. 1999, Acta Astronomica, 49, 119

\bibitem[{{Paxton} {et~al.}(2011){Paxton}, {Bildsten}, {Dotter}, {Herwig}, {Lesaffre}, \& {Timmes}}]{Paxton:11}
{Paxton}, B., {Bildsten}, L., {Dotter}, A., {et~al.} 2011, \apjs, 192, 3, \dodoi{10.1088/0067-0049/192/1/3}

\bibitem[{{Paxton} {et~al.}(2013){Paxton}, {Cantiello}, {Arras}, {Bildsten}, {Brown}, {Dotter}, {Mankovich}, {Montgomery}, {Stello}, {Timmes}, \& {Townsend}}]{Paxton:13}
{Paxton}, B., {Cantiello}, M., {Arras}, P., {et~al.} 2013, \apjs, 208, 4, \dodoi{10.1088/0067-0049/208/1/4}

\bibitem[{{Paxton} {et~al.}(2015){Paxton}, {Marchant}, {Schwab}, {Bauer}, {Bildsten}, {Cantiello}, {Dessart}, {Farmer}, {Hu}, {Langer}, {Townsend}, {Townsley}, \& {Timmes}}]{Paxton:15}
{Paxton}, B., {Marchant}, P., {Schwab}, J., {et~al.} 2015, \apjs, 220, 15, \dodoi{10.1088/0067-0049/220/1/15}

\bibitem[{{Paxton} {et~al.}(2018){Paxton}, {Schwab}, {Bauer}, {Bildsten}, {Blinnikov}, {Duffell}, {Farmer}, {Goldberg}, {Marchant}, {Sorokina}, {Thoul}, {Townsend}, \& {Timmes}}]{Paxton:18}
{Paxton}, B., {Schwab}, J., {Bauer}, E.~B., {et~al.} 2018, \apjs, 234, 34, \dodoi{10.3847/1538-4365/aaa5a8}

\bibitem[{{Paxton} {et~al.}(2019){Paxton}, {Smolec}, {Schwab}, {Gautschy}, {Bildsten}, {Cantiello}, {Dotter}, {Farmer}, {Goldberg}, {Jermyn}, {Kanbur}, {Marchant}, {Thoul}, {Townsend}, {Wolf}, {Zhang}, \& {Timmes}}]{Paxton:19}
{Paxton}, B., {Smolec}, R., {Schwab}, J., {et~al.} 2019, \apjs, 243, 10, \dodoi{10.3847/1538-4365/ab2241}

\bibitem[{{Pfahl} {et~al.}(2008){Pfahl}, {Arras}, \& {Paxton}}]{Pfahl:08}
{Pfahl}, E., {Arras}, P., \& {Paxton}, B. 2008, The Astrophysical Journal, 679, 783, \dodoi{10.1086/586878}

\bibitem[{{Rodr{\'\i}guez} \& {Breger}(2001)}]{Rodriguez:01}
{Rodr{\'\i}guez}, E., \& {Breger}, M. 2001, \aap, 366, 178, \dodoi{10.1051/0004-6361:20000205}

\bibitem[{{Rodr{\'\i}guez} {et~al.}(1995){Rodr{\'\i}guez}, {L{\'o}pez de Coca}, {Costa}, \& {Mart{\'\i}n}}]{Rodriguez:95}
{Rodr{\'\i}guez}, E., {L{\'o}pez de Coca}, P., {Costa}, V., \& {Mart{\'\i}n}, S. 1995, Astronomy and Astrophysics, 299, 108

\bibitem[{{Schenk} {et~al.}(2001){Schenk}, {Arras}, {Flanagan}, {Teukolsky}, \& {Wasserman}}]{Schenk:02}
{Schenk}, A.~K., {Arras}, P., {Flanagan}, {\'E}.~{\'E}., {Teukolsky}, S.~A., \& {Wasserman}, I. 2001, \prd, 65, 024001, \dodoi{10.1103/PhysRevD.65.024001}

\bibitem[{{Smolec} \& {Moskalik}(2008)}]{Smolec:2008}
{Smolec}, R., \& {Moskalik}, P. 2008, Acta Astronomica, 58, 193, \dodoi{10.48550/arXiv.0809.1979}

\bibitem[{Townsend {et~al.}(2018)Townsend, Goldstein, \& Zweibel}]{Townsend:18}
Townsend, R. H.~D., Goldstein, J., \& Zweibel, E.~G. 2018, Monthly Notices of the Royal Astronomical Society, 475, 879, \dodoi{10.1093/mnras/stx3142}

\bibitem[{{Townsend} \& {Teitler}(2013)}]{Townsend:13}
{Townsend}, R.~H.~D., \& {Teitler}, S.~A. 2013, \mnras, 435, 3406, \dodoi{10.1093/mnras/stt1533}

\bibitem[{{Watson}(1988)}]{Watson:1988}
{Watson}, R.~D. 1988, Astrophysics and Space Science, 140, 255, \dodoi{10.1007/BF00638984}

\bibitem[{{Weinberg} {et~al.}(2021){Weinberg}, {Arras}, \& {Pramanik}}]{Weinberg:21}
{Weinberg}, N.~N., {Arras}, P., \& {Pramanik}, D. 2021, The Astrophysical Journal, 918, 70, \dodoi{10.3847/1538-4357/ac0fdd}

\bibitem[{{Weinberg} {et~al.}(2012){Weinberg}, {Arras}, {Quataert}, \& {Burkart}}]{Weinberg:12}
{Weinberg}, N.~N., {Arras}, P., {Quataert}, E., \& {Burkart}, J. 2012, \apj, 751, 136, \dodoi{10.1088/0004-637X/751/2/136}

\bibitem[{{Weinberg} {et~al.}(2024){Weinberg}, {Davachi}, {Essick}, {Yu}, {Arras}, \& {Belland}}]{Weinberg:2024}
{Weinberg}, N.~N., {Davachi}, N., {Essick}, R., {et~al.} 2024, The Astrophysical Journal, 960, 50, \dodoi{10.3847/1538-4357/ad05c9}

\bibitem[{{Wersinger} {et~al.}(1980){Wersinger}, {Finn}, \& {Ott}}]{Wersinger:1980}
{Wersinger}, J.~M., {Finn}, J.~M., \& {Ott}, E. 1980, Physics of Fluids, 23, 1142, \dodoi{10.1063/1.863116}

\bibitem[{{Wu} \& {Goldreich}(2001)}]{Wu:01}
{Wu}, Y., \& {Goldreich}, P. 2001, \apj, 546, 469, \dodoi{10.1086/318234}

\end{thebibliography}

\end{document}